\documentclass{aa} 
\usepackage{graphicx,psfig}

\newcommand{\Msun} {M$_\odot$} 
\newcommand{\Lsun} {L$_\odot$}

\newcommand{\Lstar} {L$_\star$}

\newcommand{\um} {$\mu$m}

\newcommand{\Rrim} {R_{rim}} 
 
\newcommand{\simless}{\mathbin{\lower 3pt\hbox 
      {$\rlap{\raise 5pt\hbox{$\char'074$}}\mathchar"7218$}}} 
\newcommand{\simgreat}{\mathbin{\lower 3pt\hbox 
     {$\rlap{\raise 5pt\hbox{$\char'076$}}\mathchar"7218$}}}  
 
\begin{document}  
 
\title{Large dust grains in the inner region of circumstellar disks}

\author{
Andrea Isella \inst{1,2}, 
Leonardo Testi \inst{1}
and 
Antonella Natta  \inst{1} 
} 
\institute{ 
  Osservatorio Astrofisico di Arcetri, INAF, Largo E.Fermi 5, 
  I-50125 Firenze, Italy 
  \and 
  Dipartimento di Fisica, Universit\'a di Milano, Via Celoria 16,
  20133 Milano, Italy  
} 
 
\offprints{isella@arcetri.astro.it} 
\date{Received ...; accepted ...} 
 
\authorrunning{ISELLA, TESTI \& NATTA} 
\titlerunning{Large grains in circumstellar disks}

\abstract 
{ Simple geometrical ring models account well for
  near-infrared interferometric observations of dusty disks surrounding
  pre-main sequence stars of intermediate mass. Such models
  demonstrate that the dust distribution in these 
  disks has an inner hole and puffed-up inner edge consistent with
  theoretical expectations. }
{ In this paper, we reanalyze the available interferometric
  observations of six intermediate mass pre-main sequence stars (CQ
  Tau, VV Ser, MWC 480, MWC 758, V1295 Aql and AB Aur) in the
  framework of a more detailed physical model of the inner region of
  the dusty disk. Our aim is to verify whether the model will allow us
  to constrain the disk and dust properties.}
{ Observed visibilities from the literature are compared with
  theoretical visibilities from our model. With the
  assumption that silicates are the most refractory dust species,
  our model computes self-consistently the shape and emission of the
  inner edge of the dusty disk (and hence its visibilities for given
  interferometer configurations). The only free parameters in our
  model are the inner disk orientation and the size of the dust
  grains.}  
{ In all objects with the exception of AB Aur, our self-consistent
  models reproduce both the  interferometric results and the
  near-infrared spectral energy distribution. In four cases, grains
  larger than  $\sim$1.2\um, and possibly much larger are either
  required by or consistent with the observations. 
  The inclination of the inner disk is found to be always 
  larger than $\sim 30^{\circ}$, and in at least two objects much 
  larger.}

\keywords{}
\maketitle

\section {Introduction} 

Understanding the properties and evolution of the dust grains
contained in proto-planetary disks around pre-main sequence stars  is
important because they are the seeds from which planets may form. We
have now strong evidence that grains in disks are very different from
the grains in the diffuse interstellar medium and in the molecular
clouds from which disks  form, as reviewed, e.g., by Natta et
al. (2006). In many objects, observations with millimeter
interferometers have provided strong evidence that the grains in the
outer and cooler regions of the disk (further than 50AU from the star)
have been hugely processed, and have grown from sub-micron
sizes to millimeter and centimeter ones. Closer to the star, however,
in the regions were planets are more likely to form, observational
evidence has been confined to grains close to the disk surface. For
these, which however account for a tiny fraction of the total dust
mass, emission in the silicate features has shown a correlation
between the shape of the feature and its strength that is interpreted
as due to growth of the grains  from size $a\sim0.1\mu$m to
$a\sim1\mu$m (van Boekel et al. 2003, 2004; Meeus et al. 2003). In
this inner disk, the properties of the grains in the disk midplane are
still unknown.  

In the last few years, due to the new long baseline near-infrared
interferometers, many important steps forward in the study of the
internal regions of circumstellar disks have occurred. The available
near-infrared interferometric observations of T Tauri (TTS)
and Herbig Ae (HAe) stars (Eisner et al. 2003, 2004; Millan-Gabet et
al. 2001; Tuthill et al. 2001; Monnier et al. 2005) confirm the idea
that the inner disk properties are controlled by the dust evaporation
process which produce a ``puffed-up'' inner rim at the dust
destruction radius (Natta et al. 2001; Dullemond, Dominick \& Natta
2001, hereafter DDN01). In these models, the location and shape of the
rim depends  on the properties of grains located not on the disk
surface but on its midplane. 

Isella \& Natta (2005, hereafter IN05) have recently proposed models
of the ``puffed-up''inner rim which  include a self-consistent
description of the grain evaporation and its dependence on the gas
density. IN05 have explored a large range of grain properties, and
discussed how the location of the rim depends on grain properties. In
this paper, we will use the IN05 models to analyze the existing 
interferometric data  of the best observed HAe stars to explore, in
practice, the constraints on grain properties provided by this
technique and their uncertainties. 

As a byproduct of the modeling process, one obtains also the
orientation of the inner disk (i.e. its inclination with
respect to the line of sight and its position angle); this can be
compared with the orientation of the outer disk, obtained from
millimeter observations of the molecular gas and dust emission and/or
scattered light in the optical.

The paper is organized as follows. In \S2 we describe the available
interferometric observations of the target stars. The IN05
model for the inner rim is briefly summarized in \S3 and used
to fit the observations of the individual objects in \S4. 
A comparison of the results with  previous analysis of the same data
is presented in \S5.  Our results are discussed in \S6. Conclusions
follow in  \S7.

\section{Target stars and observations}

Our sample is composed of six HAe stars (AB Aur, CQ Tau, VV Ser, MWC
480, MWC 758 and V1295 Aql), for which near-infrared interferometric
observations  exist in the literature.  Table \ref{tab.sources}
summarizes the physical properties of the target stars. All the stars 
are classified as young stellar objects with masses ranging from 1.5
to 4.3 solar mass and a spectral type between A0/B9 and A8/F2. CQ Tau
and VV Ser belong to the family of UXORs and are characterized by
large and irregular variability.  

We use visibility measurements of the target stars from the literature,
obtained with interferometric observations carried out with PTI (Palomar
Testbed Interferometer) in K band ($\lambda_0=2.2\mu$m, $\Delta\lambda
= 0.4\mu$m) described in Eisner et al. (2004). For AB Aur and
V1295Aql, IOTA observations are also available (Millan-Gabet et
al. 2001) for the K'($\lambda_0=2.16\mu$m, $\Delta\lambda = 0.32\mu$m)
and H ($\lambda_0=1.65\mu$m, $\Delta\lambda = 0.30\mu$m) bands.     

\begin{table*}
\centering
\begin{minipage}[t]{12.5cm}
\caption{Stellar parameters.}             
\label{tab.sources}      
\begin{tabular}{c c c c l l l l}     % 8 columns 
\hline\hline       
Source & Alternate Name & Spectral Type &  d   &  T  &  L  &  M  & Av  \\
   ~   &       ~        &       ~       & (pc) &  (K) & (\Lsun) & (\Msun) & ~ \\
\hline
AB Aur    & HD 31293 & A0pe         & 144 & 9772  & 47 & 2.4 & 0.5  \\
MWC 480   & HD 31648 & A2/3ep+sh    & 140 & 8700  & 25 & 2.2 & 0.25 \\
MWC 758   & HD 36112 & A5IVe        & 230 & 8128  & 22 & 2.0 & 0.22  \\
CQ Tau	  & HD 36910  & A8 V/F2 IVea & 100 & 8000  & 5  & 1.5 & 1.00 \\
VV Ser	  & HBC 282   & B9/A0 Vevp   & 260 & 10200 & 49 & 3.0 & 3.6  \\
V1295 Aql & HD 190073 & B9/A0 Vp+sh  & 290 & 8912  & 83 & 4.3 & 0.19 \\
\hline
\end{tabular}
  \\\\
  Stellar parameters are from Hillenbrand et al. (1992), van den Ancker
  et al. (1998), Chiang et al. (2001), Strizys et al. (1996), Mannings
  et al. (2000) and references therein.
\end{minipage}
\end{table*}

\section{Model description} 
\label{sec:model}

\begin{figure*} 
  \begin{center} 
    \leavevmode 
    \centerline{ \psfig{file=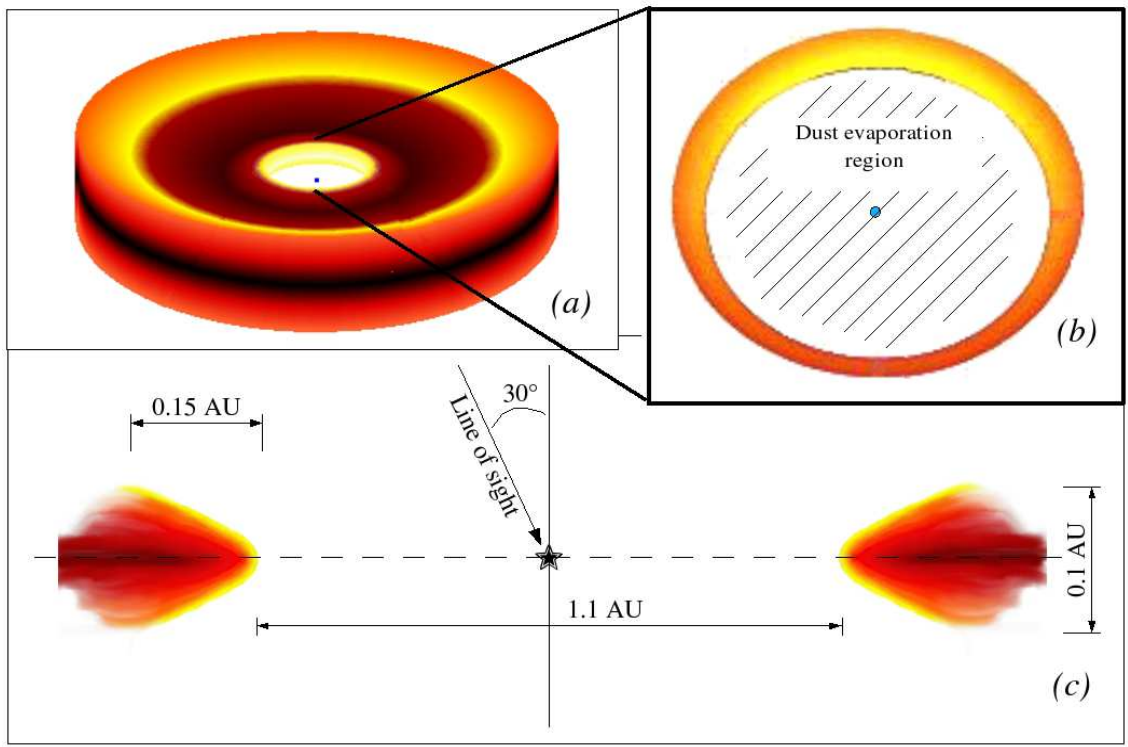,width=15cm,angle=0} } 
  \end{center} 
  \caption{Sketch of the structure of the inner part of a
  proto-planetary disk. Panel (a) shows a 
  disk with an inner puffed up rim that casts a shadow
  over the outer part of the disk. Farther than 5-10AU from the
  central star, the flaring disk emerges from the rim
  shadow. Panel  (b) presents the  image of the inner rim,
 computed with the IN05 models for a star with an effective temperature of 10000K
  and disk inclination of $30^{\circ}$. Panel (c) shows a vertical
  (edge on) section of the inner rim. The curvature of the
  surface of the rim is caused by the variation of the dust evaporation
  temperature with the height above the disk midplane (see text).
  } 
  \label{fig:disk} 
\end{figure*} 

We use a model  based on the assumption that the  near-infrared
emission of  HAe stars originates in the ``puffed up'' inner rim which
forms in the circumstellar disk at the dust evaporation radius (Natta
et al 2001; DDN01). 

In IN05 we revised the concept of the ``puffed up''inner rim,
introducing the dependence of the dust evaporation temperature on the 
local gas density, following the dust model of Pollack et
al. (1994). The main result is that the surface of the rim presents a
curved shape (see Fig.1), whose features are summarized in the 
following.

\subsection{The dust evaporation and the  ``puffed-up'' inner rim}
\label{sec:mod_1}

Following the suggestion of Natta et al. (2001), we  assume that 
the dust component of a circumstellar accreting disk is internally
truncated by  the dust evaporation process, forming an inner hole of
radius $R_{evp}$ inside of which only gas can survive. If the
radiation absorption due to this inner gas is negligible (as is 
often the case; see, e.g., Muzerolle et al. 2004), dust evaporation
occurs where the  equilibrium temperature  $T_d$ of grains  embedded
in the unattenuated stellar radiation field, equals their evaporation
temperature $T_{evp}$. In IN05 we used an analytical solution of the
radiation transfer problem (Calvet et al. 1991, 1992) to calculate the
grain temperature inside the disk and we showed that evaporation
occurs at a distance  from the star that can be expressed as: 
\begin{equation}
\label{eq:Revp}
R_{evp}[AU] = 0.034 \cdot \left( \frac{1500}{T_{evp}} \right)^2
\sqrt{ \frac{L_{\star}}{L_\odot} \left( 2+\frac{1}{\epsilon} \right) },
\end{equation}
where $L_{\star}$ is the stellar luminosity and $\epsilon$ is the
ratio of the Planck mean opacity at $T_{evp}$ to that at the
stellar effective temperature $T_\star$,
$\epsilon=\kappa_P(T_{evp})/\kappa_P(T_{\star})$.  The quantity 
$\epsilon$ measures the cooling efficiency of the grains;  it depends
on the wavelength dependence of the absorption efficiency of the
grains and varies with grain composition and size.  

If the dust in the proto-planetary disk is composed of different
types of grains, the location and  structure of the inner rim
depends on the properties of the grains with the highest evaporation
temperature. In the dust model proposed by Pollack et
al. (\cite{PH94}), the most refractory grains are silicates for which
$T_{evp}$ is given by the relation  
\begin{equation}  
\label{eq:Tevp} 
T_{evp}(r,z) = 2000 \cdot [\rho_g(r,z)]^{0.0195},  
\end{equation} 
valid for the gas density $\rho_g$ in the range between $10^{-18}g\,
cm^{-3}$ and $10^{-5}g\, cm^{-3}$.
In the following analysis, we will therefore assume that the inner
disk dust is made of silicates, with   optical properties given by
Weingartner \& Draine (2001); thus, $\epsilon$ is uniquely defined by
the grain radius $a$, and we will use $a$, rather than $\epsilon$, as
a model parameter.  

Assuming that the proto-planetary disk is in hydrostatic equilibrium
in the  gravitational field of the central star and that it is
isothermal in the vertical direction $z$, the  gas density
$\rho_g(r,z)$ has its  maximum value on the midplane  and decreases
with $z$ as  
\begin{equation}
\label{eq:rhog}
\rho_g(r,z)=\rho_{g,0}(r) \exp(-z^2/2h(r)^2) , 
\end{equation}
where $h$ is the pressure scale height of the disk. The midplane
density can be expressed as a power-law of $r$ $\rho_{g,0}(r) =
\rho_{g,0}(r_0) (r_0/r)^{\gamma}$, with $\gamma$ of the order of
2--3  (see, e.g., Chiang \& Goldreich 1997). 

The decrease of $\rho_g$ with $z$, combined with Eq.\ref{eq:Tevp},
implies that the silicate evaporation temperature varies from, i.e.,
$\sim1500K$ on the midplane (assuming a  typical gas density of
$\sim10^{-7}$g/cm$^3$) to $\sim 1000K$ at $z/h=6.4$ and $\sim 800K$ at
$z/h=8$. Since $T_{evp}$ decreases with $z$ , it is immediately clear
from Eq.\ref{eq:Revp} that the distance from the star at which dust
evaporates increases with $z$, describing a curved surface as shown in
Fig.\ref{fig:disk}.  

The dependence of $T_{evp}$ on the gas density is an important 
factor when computing the shape of the rim in the vertical direction,
where the gas density varies by many orders of magnitude while the
distance from the star is practically unchanged. In the radial
direction, we expect relatively small variations of $\rho_{g,0}$, for
any reasonable value of the disk mass, so that the distance of the rim
from the star, measured in the midplane, is practically independent of
the gas density. 

The emission of the  rim is computed assuming that it originates from
the surface characterized by an effective temperature
$T_{eff}=T(\tau_d=2/3)$, where $\tau_d$ is the optical depth for the
emitted radiation. The $T_{eff}$ surface, therefore, defines the
observed location and shape of the rim. In IN05, we discussed how the
$T_{evp}$ and the $T_{eff}$ surfaces behave for small and large
silicate grains, and showed  that the $T_{eff}$ surface moves closer
to the star  for increasing grain size until  a critical value, which
for silicates is about 1.2 \um. Larger grains produce rims with
$T_{eff}$ surfaces practically independent of $a$. Therefore, for a
fixed stellar luminosity, silicates with $a\sim 1.2$\um\ give the
minimum value of the distance of the rim from the star. Conversely, if
the measured rim distance is equal to this minimum value, one can
derive from it only a lower limit ($\sim 1.2$\um) to the grain size.

The rim emission peaks at near-infrared wavelengths. At $\lambda
\simless 5-7$\um, one can assume that the observed flux is the sum of
the stellar + rim emission, with only negligible contribution from the
outer disk  (see, e.g., DDN01). We model the  stellar photospheric
flux using standard Kurucz model atmospheres.  

\subsection{Visibility model}       
Due to the limited coverage of the $u-v$ plane of the existing
near-infrared interferometers, it is not possible at present to
recover full images from the available data, and one has to resort to
the analysis of the visibilities on given interferometric baselines. 

\begin{figure*} 
  \begin{center} 
    \leavevmode 
    %DIMENSIONE VERSIONE REFEREE
    %\centerline{ \psfig{file=fig/example.ps,width=17cm,angle=270} }
    %DIMENSIONE VERSIONE PAPER
    \centerline{ \psfig{file=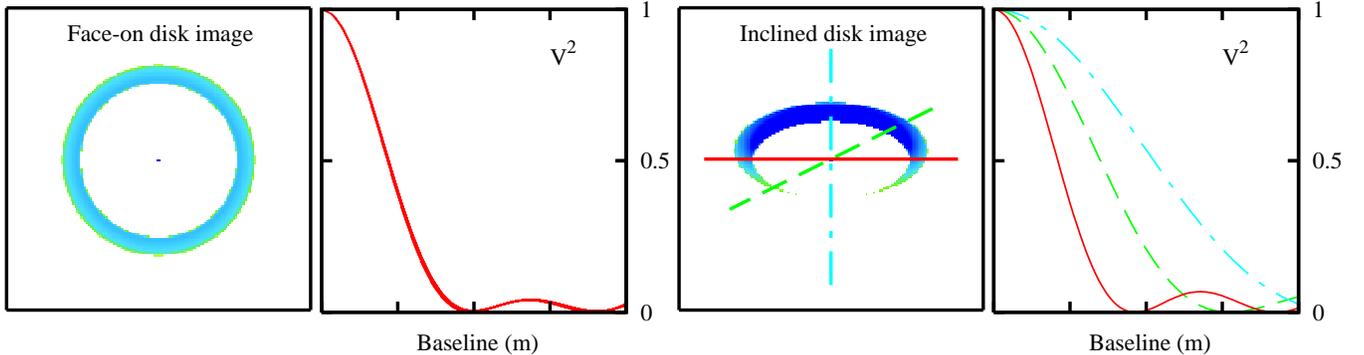,width=19.5cm,angle=270} } 
  \end{center} 
  \caption{Predicted visibilities for two different inclinations of the
  inner rim. The two panels on the left show the predicted image for a
  face-on rim  ($\iota=0^{\circ}$) and the relative visibility squared
  $V^2$: since the image is circularly symmetric, the visibility
  squared values are the same for every baseline orientation. The two
  panels on the right show the predicted image for an inclined rim
  ($\iota=60^{\circ}$). In this case, the values of $V^2$ depend on
  the baseline orientations: $V^2$ decreases rapidly for baselines
  oriented along the major axis of the image (solid line) while $V^2$
  decreases slowly for a baseline oriented along the minor axis
  (large and small dashes). For the intermediate orientation (dashed
  line), $V^2$ is in between the two extreme values.  
  } 
  \label{fig:vis} 
\end{figure*}

Starting from the synthetic images of the inner rim (see IN05), we
compute the predicted visibility values using a Fast Fourier Transform
recipe. For face-on inclination, due to the circular symmetry of the
rim image, the visibility depends only on the length of the baseline
$B$. For inclination greater than zero,  the image of the rim has an
``elliptical'' shape: the minor axis decreases with increasing inclination
and the upper half of the rim becomes brighter than the lower part.  
For baselines oriented along the direction of the minor axis of the
rim image, the visibility decreases more slowly than for those oriented
along the major axis. For all other orientations of the baseline, the
visibility will have  values intermediate between these two (see
Fig.\ref{fig:vis}). Moreover, due to the Earth rotation during the
observation, the baseline corresponding to a fixed telescope pair
moves in the u-v plane  describing an ellipse. Along this ellipse,
each point is related  to the hour angle $HA$ of the target object in
the sky. In the next section we use the $V^2$-$B$ plot and $V^2$-$HA$
plot, to show how the models fit the observations. 

The visibility model takes into account the emission of the central
star, modeled as a uniform disk of radius $R_{\star}$. If $F_{\star}$
and $F_{d}$ are respectively the stellar and the inner rim flux at the
wavelength of the observation, the total  visibility is given by the
relation: 
\begin{equation}
V^2 = \left( \frac{F_\star V_\star + F_d V_d}{F_\star + F_d}
\right) ^2,
\end{equation} 
where $V_\star$ and $V_d$ are the visibility values of the star and of 
the disk. Note that for the PTI configuration (baselines between 84m
and 100m) $V_\star$ is in all cases very closed to 1.

\section {Comparison with the observations}
\label{sec:fit} 

\subsection{Model parameters}
Once the stellar and dust properties are fixed, the model-predicted
visibilities depend on the dust grain radius $a$, which completely
defines the rim structure, and two parameters (observational parameters
in the following) that describe the orientation of the disk, namely
the inclination $\iota$ and position angle PA. 

For each star, we firstly compute the predicted rim structure varying
the size of the grains from very small to very large values. As
discussed in \S3, we assume that silicates are the most refractory
component; we take the optical properties of astronomical silicates
defined by Weingartner \& Draine (\cite{WD01}). Other disk parameters
(i.e., mass and density radial profile) can be neglected in this
analysis. We fix $\rho_{g,0}(R_{rim})\sim 10^{-7}$ g/cm$^3$ which
  gives a total disk mass of about 0.1\Msun, for a fiducial value of
  $\gamma=2.5$ and an outer disk radius of the disk of 200AU.

Once the structure of the ``puffed up'' inner rim is calculated, the
predicted  visibility depends on the orientation of the disk in
the sky, defined by the inclination $\iota$ of the midplane of the
disk with respect to the line of sight and its position angle $PA$,
measured from north to east and relative to the major axis of the
projected image of the disk on the sky. The inclination is defined
so that $\iota=0^{\circ}$ identifies a face-on disk while
$\iota=90^{\circ}$ corresponds to an edge-on disk. For inclinations
higher than 80$^\circ$ the rim emission is likely absorbed by the
outer regions of the disk and the IN05 model can not be applied. 
  
In practice, we compute  visibility models for each object
varying $a$, $\iota$ and $PA$ independently. We then select the best 
models calculating the reduced $\chi^2$ between the visibility data
and the theoretical values calculated at the same points in the u-v
plane. The observed near-infrared fluxes, and the IOTA data when
available, are then ``a posteriori'' used to check the quality
of the fit and, when possible, to reduce the degeneracy due to the
small number of visibility points. For some stars, the existing  data
do not constrain the parameters, but still define a range, outside of
which the fit to the data is very poor.    

Tab.\ref{tab:fit} shows in column 2 the best values of the
astronomical silicate radius $a$ and, in column 3, the corresponding
values of the radius of the inner rim $\Rrim$. The two observational
parameters $i$ and  $PA$ are given in columns 4 and 5,
respectively. Note that the free parameters are in boldface; $\Rrim$ 
is a derived quantity. 

\begin{table*}
\centering
\begin{minipage}[t]{17.5cm}
\caption{Best fitting model parameters. }
\label{tab:fit}      
\begin{tabular}{l|l l l l | l l l | l l } 
\hline\hline       
   ~      &   \multicolumn{4}{c}{IN05 model} &
   \multicolumn{3}{c}{Eisner et al. (2004)} & \multicolumn{2}{c}{outer 
   disk}    \\
Source    &{\boldmath $a$ } & {$\Rrim$} & {\boldmath
   $\iota$} & {\bf PA}& $\Rrim$ & $\iota$  & PA & $\iota$ & PA \\  
 ~  &($\mu$m) &   (AU)   & (deg)   &(deg) &   (AU)   &  (deg)
   &(deg) & (deg) & (deg) \\ 
\hline

MWC 758   &{\boldmath $\geq 1.2$} & 0.32 & {\bf 40} & {\bf 145} & 0.21 &
   $36^{+3}_{-2}$  & $127^{+4}_{-3}$  & 46   & $ 116^{+6 \, (a)}_{-5}$\\  

VV Ser	  & {\boldmath $\geq 1.2$} & 0.54 & {\boldmath$50-70$} &
{\boldmath$60-120$}& 0.47 & $42^{+6}_{-2}$ &  $166^{+17}_{-6}$&
$72\pm5$ & $13\pm5 ^{(b)}$ \\  

CQ Tau    & {\boldmath $0.3 - \geq 1.2$} &
  $0.16 - 0.25$ & {\boldmath $40 - 55$} & {\boldmath$145-
  190$} & 0.23& $48^{+3}_{-4}$ & $ 106^{+4}_{-5}$
  &$63^{+10}_{-15}$  &  $2\pm13^{(c)}$   \\  

V1295 Aql & {\boldmath $0.3 - \geq1.2$} & $0.7 -
  1.2$  & {\boldmath$40 - 65$}  &  &0.55 &
$23^{+15}_{-23}$ &  & ~ &\\ 

MWC 480   & {\boldmath $0.2 - 0.3$} & $0.53 
  - 0.63$ & {\boldmath$30 - 65$} & ~& 0.23 & $28^{+2}_{-1}$
& $145^{+9}_{-6}$  & $ 20 - 40$ &$147 - 180^{(a,d)}$ \\ 

AB Aur    & \multicolumn{3}{c}{impossible to fit} && 0.25 &
$8^{+7}_{-8}$ &  &$15 - 35$ &$50 - 110 ^{(e,f,g,h)}$  \\ 

\hline
\end{tabular}
\\\\
 From column 2 to 5 are reported the best fit parameters for the
 ``puffed-up'' inner rim, obtained with the IN05 model: the grain
 radius $a$, the radius of the inner rim $\Rrim$, the inclination
 $\iota$ and the position angle $PA$. The free parameters of the model
 are presented in bold face. Columns 6,7 and 8 show the values of the
 radius of the inner rim, the inclination and the position angle,
 obtained by Eisner et al. (2004). Finally, the last two columns show
 the available estimates of inclination and position angle for the 
 external region of the disk: $(a)$ Mannings et al. (1997); $(b)$
 Pontoppidan et al. (2006); $(c)$ Testi et al. (2001, 2003); $(d)$
 Simon et al. (2000); $(e)$  Fukagawa et al.(2004); $(f)$ Grady et
 al. (1999); $(g)$ Corder et al. 2005; $(h)$ Pi\'etu et al. (2005).  
\end{minipage}
\end{table*}

\subsection{MWC 758}
\label{sec:MWC758}
\begin{figure*} 
  \begin{center} 
    \leavevmode 
    % DIMENSIONE VERSIONE REFEREE
    %\centerline{ \psfig{file=fig/MWC758_vis2+sed.ps,width=16cm,angle=270}  }
    % DIMENSIONE VERSIONE PAPER
     \centerline{ \psfig{file=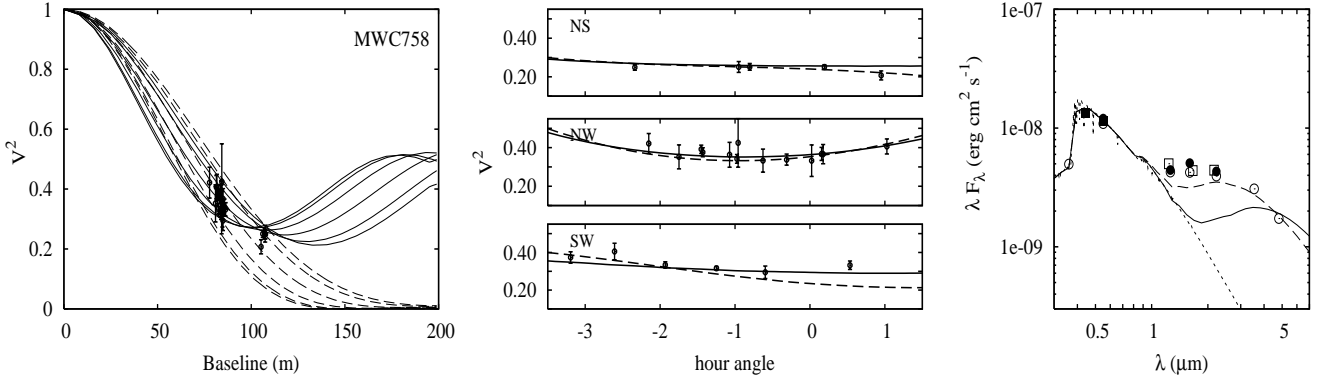,height=5.2cm,width=18cm,angle=270} } 
  \end{center} 
  \caption{ The left and central panels show K-band $V^2$ data for MWC
  758 respectively plotted as function of the baseline
  and of the hour angle, for the three different PTI baseline
  orientations (NS, baseline legth of 110m in direction North-South;
  NW, 86m direction North-West; SW, 87m direction South-West). 
  PTI measurements (Eisner et al. 2004) are shown by dots. The right
  panel shows the spectral energy distribution of MWC 758. The
  de-reddened photometric fluxes are from Eisner et al. (2004, filled
  circles), Malfait et al. (1998, empty circles), van den Ancker et
  al. (1998,  filled squares) and Cutri et al. (2003, empty squares); 
  the dotted line shows the photospheric stellar flux.
  The solid lines show the predictions of the best-fitting model with
  small grains ($a=0.17\mu$m, $\iota=48^{\circ}$, $PA=134^{\circ}$);
  the dashed lines the best-fitting model with big grains
  ($a\geq1.2\mu$m, $\iota=40^{\circ}$, $PA=145^{\circ}$). As described
  in \S\ref{sec:MWC758}, all the intermediate disk configuration can
  reproduce the observations at almost the same level of
  confidence. Each curve on the left panel corresponds to a different
  orientation of the baseline on the plane of the sky (see
  Fig.\ref{fig:vis}). The stellar parameters are given in Table 1.   
  }
  \label{fig:MWC758_vis2} 
\end{figure*}

PTI visibilities are fitted by a family of models, with parameters
varying between the two extreme cases shown in
Fig.\ref{fig:MWC758_vis2}. In one case, the disk has small grains of
radius $a=0.17$ $\mu$m, $\iota= 48^{\circ}$ and $PA=134^{\circ}$; in
the other, big grains with $a \geq 1.2 \mu$m, $\iota =40^{\circ}$ and
$PA= 145^{\circ}$. Models with $a$ values within this range
will fit the observed visibilities equally well, provided that we vary
$\iota$ and $PA$ in an appropriately way.

However, if we consider also the constraints set by the SED at
near-infrared wavelengths, we find that only models with big grains
fit it reasonably well (see the right panel of Fig.\ref{fig:MWC758_vis2}). The
best-fitting model $(\chi^2_r=2.0)$ has then $a \geq 1.2$\um, inner rim radius 
is $\Rrim=0.32$AU, rim effective temperature  (at $z$=0) is 1460K. The
near-infrared flux, $L_{NIR}$, integrated between 2$\mu$m and 7$\mu$m
is $25\%$ of the total stellar luminosity, similar to the observed
value. Once we fix $a$, the formal uncertainties on $\iota$,
estimated from the surface where the reduced $\chi^2$ equals
$\chi^2_{min}+1$, are quite small, $\pm 3^{\circ}$. More realistic
uncertainties are of the order of $10^{\circ}$ for both $\iota$ and
$PA$.  

Note that in MWC 758 the PTI visibilities define quite well the
orientation of the disk, even when the SED is not
used to constrain the grain size. In particular, the inclination
cannot be lower than about $30^\circ$.

\subsection{VV Ser}
\begin{figure*} 
  \begin{center} 
    \leavevmode 
    % DIMENSIONE VERSIONE REFEREE
    %\centerline{ \psfig{file=fig/VVSer_vis2.ps,width=16cm,angle=270} }
    % DIMENSIONE VERSIONE PAPER
    \centerline{ \psfig{file=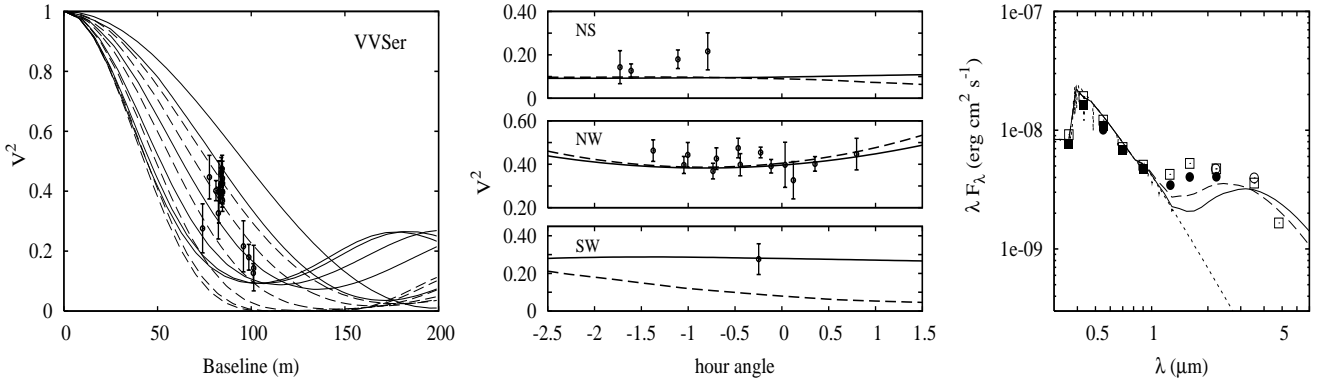,height=5.2cm,width=18cm,angle=270} } 
  \end{center} 
  \caption{Same as Fig.\ref{fig:MWC758_vis2} for VV Ser.
  The solid curves show the model predicted values of the K-band $V^2$ for a rim
  with $a=0.52\mu$m, $\iota=65^{\circ}$ and $PA=80^{\circ}$; the
  dashed curves  for a rim model with $a\geq1.2\mu$m, $\iota=55^{\circ}$ and
  $PA=115^{\circ}$. The photometric data are from Eisner et al. (2004,
  filled circles), Cutri et al. (2003, empty squares), Hillenbrand et
  al. (1992,  empty squares), Rostopchina et al. (2001, filled
  squares). Stellar parameters in Table 1.
  }
  \label{fig:VVSer_HA} 
\end{figure*}

The  results for the star VV Ser are shown in
Fig.\ref{fig:VVSer_HA}. As for MWC 758, the interferometric
observations allow  different sets of model parameters.
Namely, we obtain similar values of the reduced $\chi^2$
$(\sim1.2)$ for all grain sizes $a\simgreat 0.4$\um. Over this
range of $a$, inclination and position angle vary in the  
intervals  $45^\circ - 80^\circ$ and $60^\circ - 120^\circ$,
respectively, with lower inclinations  for larger grains.
The correlation between $\iota$ and $PA$ is very strong,  and
the uncertainties in these two parameters remain very large even for
fixed $a$. 

Although the fit is never very good, the VV Ser SED is better
accounted for by large grains (see the right panel of
Fig.\ref{fig:VVSer_HA}) and in Table 2 we show the best values of the
parameters for $a\geq 1.2$\um. The rim effective temperature is 1400
K, the near-infrared excess is 21\% of \Lstar.  

\subsection{CQ Tau}
\begin{figure*} 
  \begin{center} 
    \leavevmode 
    % DIMENSIONE VERSIONE REFEREE
    %\centerline{ \psfig{file=fig/CQTau_vis2.ps,width=16cm,angle=270} } 
    % DIMENSIONE VERSIONE PAPER
    \centerline{ \psfig{file=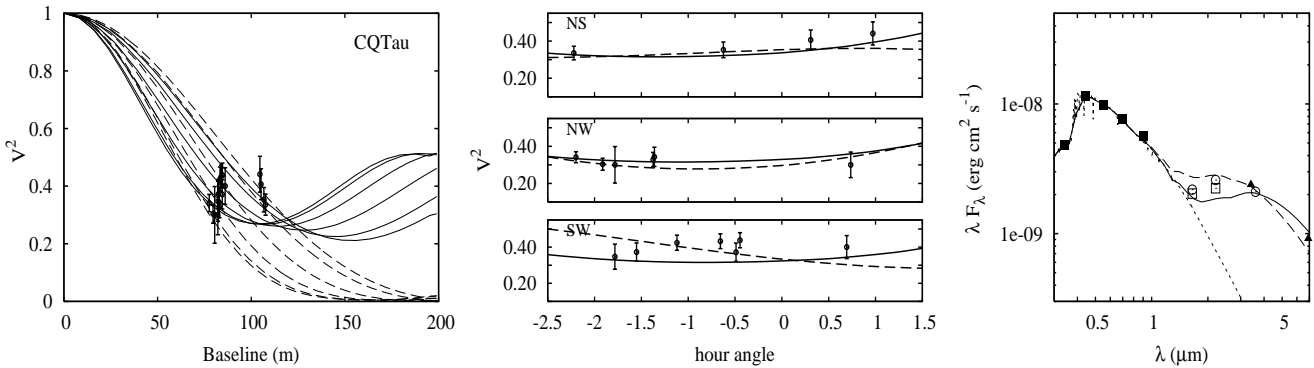,height=5cm,width=18cm,angle=270} } 
  \end{center} 
  \caption{Same as Fig.\ref{fig:MWC758_vis2} for CQ Tau.
    The solid curves show the predictions of the model with
    $a=0.3\mu$m ($\Rrim=0.25$AU), $\iota=52^{\circ}$ and
    $PA=168^{\circ}$; the dashed curves the model with $a\geq1.2\mu$m 
    ($\Rrim=0.16$AU), $\iota=46^{\circ}$ and $PA=164^{\circ}$. The
    photometric data are from Eisner et al. (2004, empty squared), Glass \&
    Penston (1974, filled squares), ISO catalogue (empty circles) and
    Natta et al. (2001, filled triangles) and references therein.} 
  \label{fig:CQTau_HA} 
\end{figure*}

The limited number of visibility points does not allow us to constrain
all the parameters of the disk. Fig.\ref{fig:CQTau_HA} shows two
models, with the same level of confidence $(\chi^2_r \sim 1)$;   
the two models have similar orientations (inclinations of 52$^{\circ}$
and 46$^{\circ}$ with position angles of 168$^{\circ}$ and
164$^{\circ}$ respectively) but very different grain sizes
($a=0.3\mu$m and $a \geq 1.2 \mu$m) and radii of the inner rim (0.25AU and
0.16AU, respectively). For $a \geq 1.2 \mu$m, the effective
temperature of the rim (at z=0) is $T_{eff}=1480K$, while
$T_{eff}=1050$ for $a=0.3\mu$m. The  SEDs of the two models are
compatible with the observed fluxes, with $L_{NIR}=13\%-18\%$
\Lstar. All the models with  $a$ within this range, and similar
$\iota$ and $PA$, have similar $\chi^2$ values. Outside this range,
models give a much poorer fit to the data. 

As for MWC 758, the visibility data constrain well the orientation of the
disk on the sky. The inclination, in particular, has to be quite
large, $40^{\circ} \simless \iota \simless 55^{\circ}$.

\subsection{V1295 Aql}
\label{sec:V1295Aql}
The PTI observations of V1295 Aql are characterized by a very small number
of visibility points and the disk parameters are hardly constrained. Even
adding the IOTA data does not help due to the big
errors that affect these observations. Note also that  PTI and IOTA
observations are performed at different wavelengths, K and H respectively.

\begin{figure*} 
  \begin{center} 
    \leavevmode 
    % DIMENSIONE VERSIONE REFEREE
    %\centerline{ \psfig{file=fig/V1295Aql_vis2.ps,width=16cm,angle=270} } 
    % DIMENSIONE VERSIONE PAPER
    \centerline{ \psfig{file=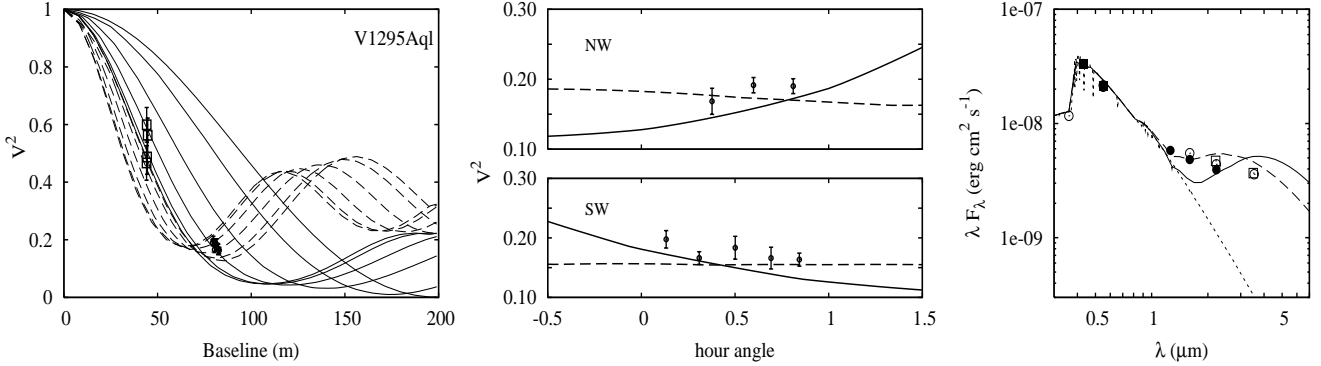,height=5cm,width=18cm,angle=270} } 
  \end{center} 
  \caption{Same as Fig.\ref{fig:MWC758_vis2} for V1295 Aql. The solid
  lines plots the results of a model with $a\geq1.2\mu$m
  ($\Rrim=0.7AU$),  $\iota=63^\circ$ and $PA=162^\circ$. The dashed
  lines a  model with $a=0.3\mu$m ($\Rrim=1.2$AU), $\iota=40^\circ$ and $PA=80^\circ$.
  The photometric measures are from Eisner et al. (2004, filled circles),
    Malfait et al. (1998, open circles), Kilkenny et al. (1985, filled
  squared) and Glass \& Penston (1974, open squared). The short
  baseline points in the left panel (empty squares) are from IOTA,
  while the central panel shows only the PTI data. 
  } 
  \label{fig:V1295_HA} 
\end{figure*}

As for CQ Tau, we show the models for the two extreme sets of
parameters that give an equally good fit (Fig.\ref{fig:V1295_HA})
with $\chi^2_r \sim 1$. All the intermediate combinations of $a$,
$\iota$ and PA can explain the observations as well. In order to fit
the values of visibility, an inclination ranging between $40^\circ$
and $65^\circ$ is required. More face-on systems can not in general
reproduce the visibility spread in the IOTA data and the $V^2-HA$
behaviour of the PTI points. The grain radius varies from
$a\geq1.2\mu$m ($\Rrim=0.7$AU, $T_{eff}=1400K$) to $a=0.3\mu$m
($\Rrim=1.2$AU, $T_{eff}=970K$) while the position angle  can not be
defined at all.   

The right panel of Fig.\ref{fig:V1295_HA} shows the comparison between
the observed and the predicted SED. More inclined disks can in general
reproduce better the photometric measurements around 1.5$\mu$m. The
near-infrared emission of disk with an inclination of less than
50$^\circ$ is peaked at about $4\mu$m and  can not reproduce the
infrared excess between 1$\mu$m and 2$\mu$m; in both models, the
integrated near-infrared flux  is $L_{NIR}\sim 20\%$ \Lstar. 

\subsection{MWC 480}
\begin{figure*} 
  \begin{center} 
    \leavevmode 
    % DIMENSIONE VERSIONE REFEREE
    %\centerline{ \psfig{file=fig/MWC480_vis2.ps,width=16cm,angle=270} } 
    % DIMENSIONE VERSIONE PAPER
    \centerline{ \psfig{file=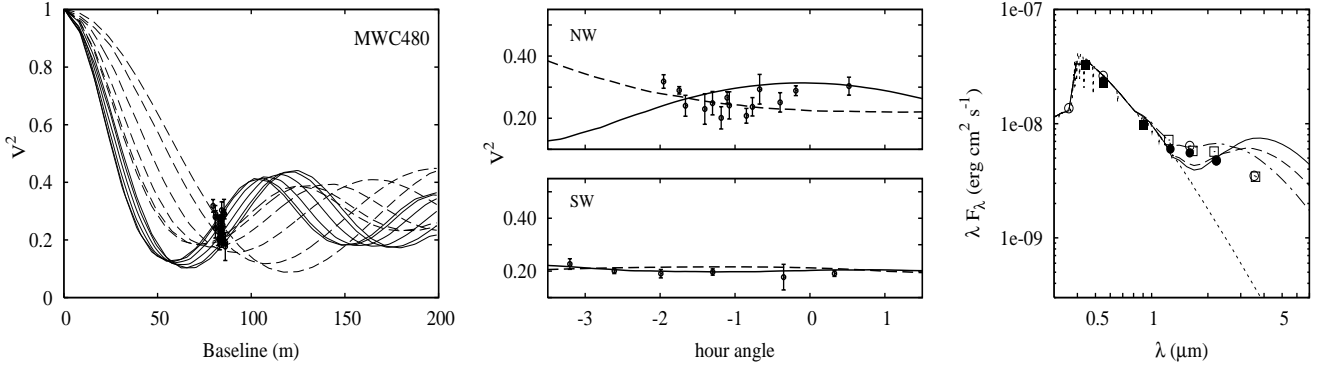,height=5cm,width=18cm,angle=270} } 
  \end{center} 
  \caption{Same as Fig.\ref{fig:MWC758_vis2} for MWC 480. 
  The solid lines are for the disk model characterized by
  $a=0.2\mu$m ($\Rrim = 0.63$AU), $\iota=35^{\circ}$ and
  $PA=30^{\circ}$. The dashed lines are relative to a disk model
  characterized by $a=0.3\mu$m ($\Rrim=0.53$), $\iota=60^{\circ}$ and
  $PA=102^{\circ}$. 
  The photometric data are from Eisner et al. (2004, filled
  circles), Malfait et al. (1998, open circles), Cutri et al. (2003,
  empty squares) and van den Ancker et al. (1998, filled squares). } 
  \label{fig:MWC480_vis2} 
\end{figure*}

Also in this case, due to the narrow range of available baselines, the
PTI visibilities of MWC 480 are consistent with different sets of
parameters at the same level of confidence $(\chi^2_r\sim2)$. 
Fig.\ref{fig:MWC480_vis2} shows the two extreme disk configurations
characterized by quite similar parameters for the dust grain size
($a=0.2\mu$m$-0.3\mu$m; $\Rrim=0.63$AU-0.53AU and  $T_{eff}\simeq
1250K$), but very different values of the inclination
($\iota=35^{\circ}$ and $\iota=60^{\circ}$) and the position angle
($\psi=60^{\circ}$ and $\psi=168^{\circ}$). Several intermediate
configurations reproduce the observed data as well. The degeneracy can 
not be removed even using the SED (Fig.\ref{fig:MWC480_vis2}) which is 
similar in the two models ($L_{NIR}/L_{\star} = 0.18\%-0.14\%$) and it
is only roughly consistent with the photometric values.  

\subsection{AB Aur}
\label{sec:fitABAur}
\begin{figure} 
  \begin{center} 
    \leavevmode 
    \centerline{ \psfig{file=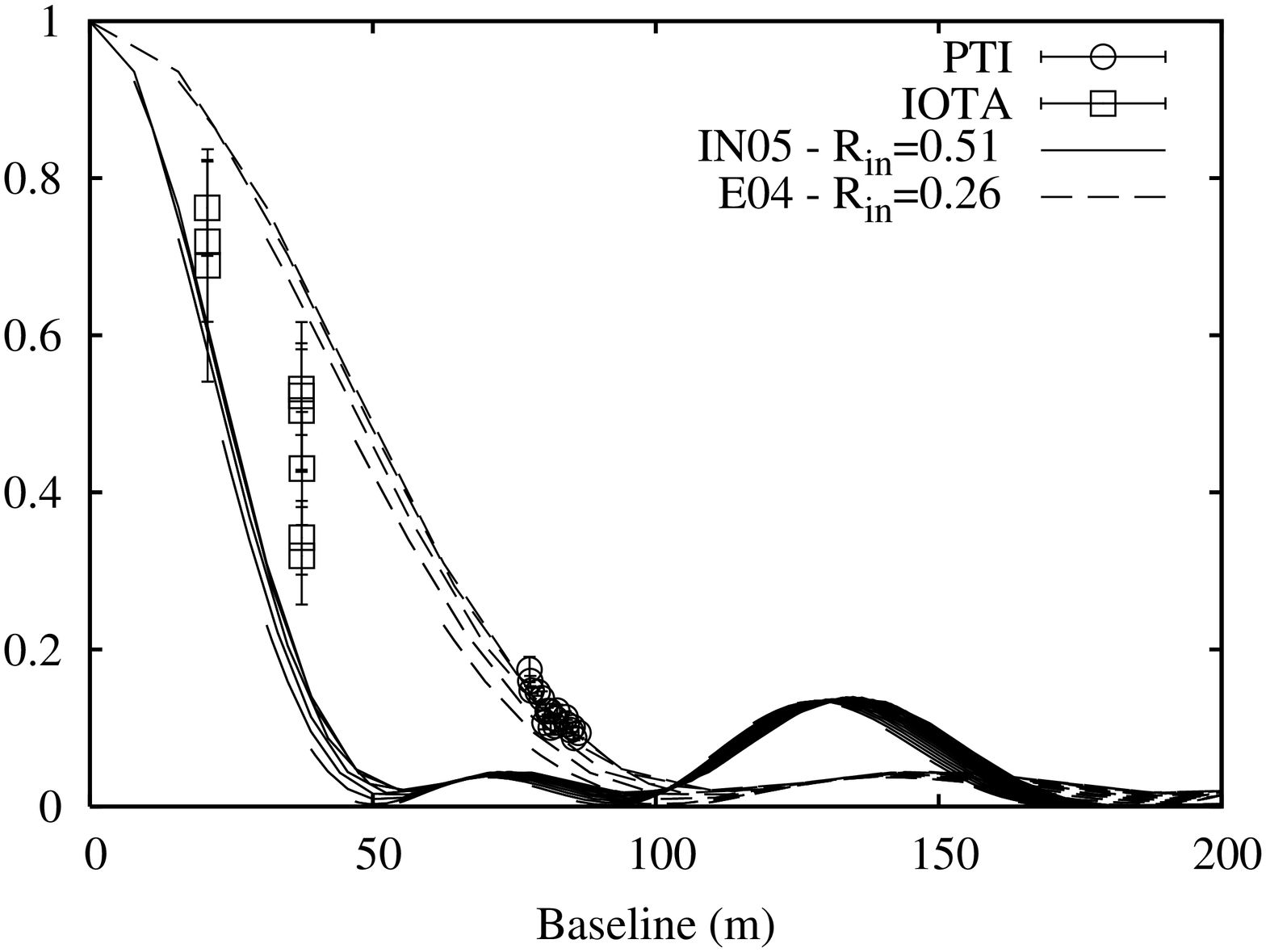,width=9cm,angle=270} } 
  \end{center} 
  \caption{$V^2$ data for AB Aur plotted in function of the
  baseline. The square and circle points refer respectively to the
  IOTA and PTI observations in K band. The solid line is relative to
  the best prediction of the self-consistent model of inner rim
  ($a\geq1.2\mu$m, $\Rrim=0.51$AU, $\iota=20^{\circ}$). The dashed
  line is obtained with the IN05 model using $a\geq1.2\mu$m with an
  {\it ad hoc} $T_{evp}=2800$K ($\Rrim=0.26$AU).  
  }
  \label{fig:ABAur_vis2} 
\end{figure}

AB Aur is the only HAe star that cannot be fitted with the IN05
models. The PTI visibilities require a face-on rim, consistent with
the inclination derived from large-scale images in scattered light
(Grady et al.~1999; Fukagawa et al.~2004) and at millimeter
wavelengths (Corder et al.~2005; Pi\'etu et al.~2005). For these
inclinations, the $V^2$ data imply a very small inner radius, about
two times smaller that the smallest $R_{rim}$ obtained using the IN05
model (see Fig.\ref{fig:ABAur_vis2}). 
If, to put the discrepancy in a more physical context, we take
$T_{evp}$ as a free parameter, we find good agreement with the PTI
data for $T_{evp}\sim 2800K$ (dashed line), a value by far too high
not only for silicates but also for any other type of grains (e.g.,
Pollack et al.~1994). 

The situation becames even less clear if we consider also the IOTA
observations (squared points), since they seem to indicate the
presence of a more inclined disk with an inner radius  between 0.26AU
and 0.51AU. No additional information can be obtained from the
analysis of the spectral energy distribution, since all the inner rim
models with an effective temperature between 1500K and 2500K are
compatible with the photometric data. We will come back to AB Aur in
\S 6. 

\section {Comparison with previous analysis}
Fits to the same interferometric data analyzed in \S\ref{sec:fit} have
been obtained by Eisner et al. (2004, hereafter E04) assuming a
toroidal shape for the inner ``puffed-up'' rim, based on the
simplified DDN01 model. In these fits, the free parameters are the
location of the rim $\Rrim$ and the two observational parameters,
$\iota$ and $PA$. The E04 results are shown in Table 2. 

We note that for three objects (MWC 758, VV Ser and CQ Tau) the E04
inclinations are in agreement within the errors with the values
obtained with the IN05 model, while for the other two (V1295 Aql and
MWC 480) the E04 $\iota$ estimates are consistent with the lowest
value of the range derived in this paper. 

The largest differences  are in the derived values of $R_{rim}$:  the
IN05 inner radii are always larger than E04 results, with a maximum
difference of a factor $\sim 3$ if we consider our maximum $\Rrim$ in
the MWC 480 system. While for CQ Tau the two values are almost the 
same, for all the other stars the difference is a factor 1.5 and 2.
This discrepancy is  mainly due to the difference between IN05 and E04
models. In particular, in IN05, the curved shape of the emitting
surface is self-consistently calculated, allowing a more correct
determination of the dependence of the rim emission on the inclination
of the disk.  

Moreover, the IN05 model takes into account the effect of the
radiation transport within the disk, even if in an approximate way (see
Appendix A in IN05). This supplementary heating is neglected
by E04, who calculate the dust temperature taking into account only
the direct stellar radiation. The ratio between the two values of
$\Rrim$ is given by the relation 
\begin{equation}
\label{eq:R/R}
\frac{ \Rrim }{ \hat{R}_{rim} } = \sqrt{\hat{\epsilon} \left(2 +
  1/\epsilon  \right) },  
\end{equation}
where the $\hat\epsilon$, $\hat{R}_{rim}$ and $\hat{T}_{evp}^2$ are the
values used by E04 and the inner radius $\hat{R}_{rim}$ is given by the
relation 
\begin{equation}
\label{eq:R04}
\hat{R}_{rim} = \frac{1}{\hat{T}_{evp}^2} \sqrt{ \frac{L_{\star}}{4\pi\sigma}
   {\frac{1}{\hat{\epsilon} }  } }.
\end{equation}
Assuming the same value of the dust emissivity 
($\epsilon=\hat{\epsilon}$), the ratio $\Rrim/\hat{R}_{in}$ is $\sim1$
for $\epsilon<<1$. The difference increases for larger $\epsilon$ and
is maximum when $\epsilon$ and $\hat\epsilon$ are very different.  

Finally, there also differences due to the fact that E04 assumes
$T_{evp}$ (or, equivalently, $\Rrim$) as a free parameter, while in 
the IN05 model $T_{evp}$ is self-consistently determined starting from 
the choice of the type of grains and the gas density in the disk (see 
Eq.\ref{eq:Tevp}). For all our target stars, the resulting values of 
$T_{evp}$ vary between 1370K and 1460K, and are  in some cases 
significantly different from those given by E04.

\section{Discussion} 

The results presented in \S\ref{sec:fit} show that, with the exception
of AB Aur, the IN05 self-consistent models of the ``puffed-up'' inner
rim can explain the available observations, both visibilities and
SEDs, of HAe stars. They can be used to derive information
about the properties of the dust present in the innermost region of
the circumstellar disk, the location of the inner rim and the
orientation of the disk on the sky, using a minimum number of
assumptions. 

\subsection{Presence of large grains}
\label{sec:large}
As shown in \S\ref{sec:fit}, the IN05 models reproduce the
interferometric data under the assumption that the most refractory
dust in the inner disk is made of silicates, with properties typical
of astronomical silicates (Weingartner \& Draine \cite{WD01}). 

In four cases, grain sizes larger than $\sim 1.2$\um\ are either
required by or consistent with the observations. Only in one
case are the data better fitted with $a \sim
0.2-0.3\mu$m. Grains in the rim  are thus larger, and often much
larger, than  grains  in the interstellar medium ($a=0.01-0.1\mu$m,
Weingartner \& Draine \cite{WD01}), confirming that grain growth has
taken place in the innermost disk regions (van Boekel et
  al. 2004).

Even if the predicted near-infrared excess
agrees well with the photometric observations, some interesting 
differences exist between the theoretical and the observed spectral
energy distributions. With the exception of CQTau, the predicted SEDs
always peak at a wavelength slightly longer than found in the
observations: the flux at short wavelengths (between 1.5\um\
and 2.2\um) is thus generally underestimated while the flux
between 2.2\um\ and 7\um\ is overestimated. This may be due to the
fact that in our models the SED is computed assuming that each point
on the surface of the rim emits as a black body at the local effective
temperature. This approximation is energetically correct but may not
reproduce the exact wavelength dependence of the emitted radiation
(see Appendix in IN05). 

The rim models fail only in the case of AB Aur, where silicates,
of whatever size, produce rims that are too distant from  the star to
be consistent with the observations. As shown in \S 4.7, PTI and IOTA
data give somewhat contradictory results, and more interferometric
data are clearly required. However, unless further observations
  drastically change  the present picture, the discrepancy between
the rim model predictions  and the data is  highly significant, and
some of the basic underlying assumptions need to be changed. It is
possible that in AB Aur grains more refractory than silicates dominate
the dust population in the inner disk; however, the $T_{evp}$ required
($\sim 2800$K)  is too high for any dust species likely present in
disks (e.g., Pollack et al. 1994). 

It is more likely that gas in the dust-depleted inner region absorbs a
significant fraction of the stellar radiation, shielding the dust 
  grains which are therefore cooler than in our models. This
  requires a high gas density in the inner disk, as expected  if the
  accretion rate is high;  in general, the accretion rates of HAe
stars (including AB Aur) are low enough to ensure that the gaseous
disk  remains optically thin (Muzerolle et al.~2004). However, our
knowledge of the accretion properties and gas disks of HAe stars is
very poor, and should be investigated further.  

AB Aur may be more than just an oddity. The presence of optically
thick gas inside the inner rim has been proposed to explain the
near-infrared interferometric observations of some very bright
Herbig Be stars, for which the visibility data  suggest inner disk
radii many times too small to be consistent with the ``puffed-up'' rim
models (Malbet et al. 2005, Monnier et al. 2005). If this is the case,
AB Aur could be the low-luminosity tail of the same phenomenon, which,
given its small distance and large brightness, could be used to
understand a whole class of objects.

\subsection{Inclination and position angle}

Inclination and $PA$ are well constrained by the existing data only in
one case (MWC~758). We want to stress,  however, that values of the
parameters outside  the ranges given in Table 2 do not fit the data at
all. In particular, there are no objects consistent with face-on
disks, or in general with a centro-symmetric brightness
distribution. This rules out models, such as those of Vinkovic et
al. (2005), where most of the near-infrared flux is contributed by a
spherically symmetric shell around the star, rather than by a
circumstellar disk. 

Two stars (CQ Tau and VV Ser) belong to the group of UXOR
variables, which are interpreted as objects with disks seen close to
edge-on (Grinin et al. 2001, Natta \& Whitney 2001, Dullemond et
al. 2003). We derive for them large inclinations,  in agreement with
this interpretation. 
 
For some of our targets, there are in the literature estimates of the
orientation of the outer disk  on the plane of the sky obtained
with millimeter interferometers (Testi el al. 2001, 2003; Manning \& Sargent
1997; Pi\'etu et al.~2005; Corder et al.~2005). These determinations
refer to the outer disk, i.e., to spatial scales of 50--100 AU at
least. The comparison with the values  derived in the
near-infrared for the inner disk (on scales of less than 1 AU) can
provide information on possible distortions in the disk,
e.g. variations of the inclination with radius.
For the three disks with millimeter data (MWC 758, CQ Tau and MWC 480),
there is agreement (within the uncertainties) between the inclination
obtained by  infrared and  millimeter observations 
However, it is certainly premature to exclude the existence of
disk distortions, given the large uncertainties that affect both the
millimeter and the near-infrared estimates. More accurate
interferometric observations in the two wavelength ranges and
self-consistent models of the  disk at all physical scales are
required.    

An interesting case is that of VV Ser, whose disk has recently
been imaged as a shadow seen against the background emission in the
11.3 PAH feature (Pontoppidan et al. 2006); These authors derive an
inclination (of the outer disk) of about 70$^{\circ}$ and a position
angle of $13^\circ \pm 5^\circ$. While the inclination is consistent
with the upper limit of the range we obtain for the inner disk,  the
position angle is off by almost $90^\circ$. This discrepancy is
intriguing, and deserves further investigation.

\subsection{Improving the model constrains}  
\label{sec:Const} 

Near-infrared interferometric observations of disks around pre-main
sequence stars are still few and sparse. It is clear from our analysis
that even in the most favorable cases more visibility data at
different baselines are necessary to narrow the range of possible disk
inclinations and grain properties. 

Given the huge demands of telescope time that interferometric
observations require, it is useful to make use of model
predictions in preparing the observations and in choosing the baseline
configurations that can constrain the disk structure. 

CQ Tau represents an example of how the IN05 model can be used in this
context. To better constrain the inner rim structure, one  will
need  observations with baselines longer than 130m (available
with the VLT interferometer), for which the predicted values of the
squared visibility parameters are very different for different disk
models (see Fig.\ref{fig:CQTau_HA}). On
the other hand, observations with baseline shorter than 60m could
better constrain the inner radius of MWC 480, since a degeneracy in
the models is present at longer baselines. 

V1295 Aql represents a still different case, in which the degeneracy
in the values of the predicted squared visibility can be removed
observing at distant hour angles for the same baseline configurations,
in order to determine the visibility variations at the same baseline,
due to the inclination of the disk.

\section{Summary and conclusions}

In this paper we have analyzed the near-infrared interferometric
observations of the six best observed HAe stars using the rim models
developed by Isella \& Natta (2005). Our  aim was to explore the
potential of near-infrared interferometry to constrain the properties
of the grains in the inner disks of these stars. 

The basic assumptions of the IN05 rim models are that the inner
disk structure is controlled by the evaporation of dust in the
unattenuated stellar radiation field, as expected if the gaseous disks
have low optical depth, and that the most refractory grains are
silicates. The IN05 self-consistent models for the ``puffed-up''
  inner rim reproduce both the interferometric observations and the
near-infrared spectral energy distribution of all the objects we have
studied, with the exception of AB Aur, which we have briefly 
discussed.

For the five stars where we are able to obtain a good fit to the data,
we can estimate the grain sizes in the rim, i.e., in the midplane of
the inner disk. We find that in  four cases grains larger than $\sim
1.2$\um\ are either required by or consistent with the
  data. Only in one case do we find that the existing data require
$a\sim 0.2-0.3$\um. Note that  this value of $a=1.2$\um\  is  a lower
limit to the grain size: grains can be much larger, since the
rim location and shape do not change significantly if the grains grow
further.  

As a result of the model-fitting, one derives also the inclination and
position angle of the disk on the plane of the sky. We find that, in
general, these parameters are not well constrained by the
  existing data. However, in all cases we can fit, inclinations lower than
$30^{\circ}$ are not consistent with the observations and the surface
brightness distribution can not be circularly symmetric.  This rules
out a spherical envelope as the dominant source of the
near-infrared emission.  

For some objects, estimates of the inclination of the outer disk  have
been obtained from millimeter interferometric observations; within the
uncertainties, they agree with the values obtained for the inner disk. 

Our analysis shows that near-infrared interferometry is a very
powerful tool for understanding the properties of the inner disks, in
particular when combined with physical models of these
regions. However, at present the existing  data are for many objects
still too sparse in their coverage of the u-v plane to allow an
accurate determination of  the disk parameters. We expect that this
will be improved in the future.  In this context, since near-infrared
interferometry is and will remain a very time demanding  technique, we
stress the importance of using physical models of the inner
region of the disk in planning future observations. 

\begin{acknowledgements}
We are indebted to Josh Eisner and Rafael Millan-Gabet for
providing us with the PTI and IOTA data. The authors acknowledge partial
support for this project by  MIUR PRIN grant 2003/027003-001.

\end{acknowledgements}

\end{document}